\documentclass[aps]{revtex4}

\include{psfig}  
\begin{document}
\title{\bf Measurements of surface impedance of superconductors as a 
function of frequency in microwave range}
\author{S.Sarti}
\affiliation{Dipartimento di Fisica and  INFM, Universit\`a 
di Roma {\it La Sapienza},\\
 P.le Aldo Moro 2, I-00185, Roma, Italy}
\author{C.Amabile}
\affiliation{Dipartimento di Fisica and  INFM, Universit\`a 
di Roma {\it La Sapienza},\\
 P.le Aldo Moro 2, I-00185, Roma, Italy}
\author{E.Silva}
\affiliation{Dipartimento di Fisica ``E.Amaldi'' and 
INFM,\\ Universit\`{a} ``Roma Tre'', Via della Vasca Navale 84, 00146
Roma,Italy}

\date{\today} 

\begin{abstract}
We report measurements of the complex resistivity in YBCO and MgB$_2$ thin films over a continuous frequency spectrum in the microwave range, making use of a Corbino disk geometry. The paper mainly focuses on the extraction of the resistivity from raw data, displaying data analysis procedure and its limits of validity. We obtain and show resistivity curves as a function of frequency and temperature denoting a frequency dependent widening of the superconducting transition.
\end{abstract}

\maketitle
\section{Introduction}

The microwave response of superconductors is of great interest in several  temperature and magnetic field ranges, since it carries information about both macroscopic and microscopic quantities, such as resistivity or electromagnetic penetration depth.

Most of available experimental data are restricted to fixed frequencies (resonant cavity techniques), whereas a study over a continuous frequency spectrum would allow for a deeper investigation of the dynamical behaviour of the system. An experimental technique which makes it possible to evaluate the 
complex surface impedance $Z_{s}=R_{s}+iX_{s}$ of a superconducting film as a function of frequency has been proposed in the past by some groups \cite{anlage,toso1}. The technique is based on the measurement of the reflection coefficients at the input of a coaxial line terminated by the film (Corbino disk). 

Extracting the contribution of the sample from the measured data is a critical step during data analysis: the contribution of the coaxial line, in fact, is in general comparable with that of the sample. The main purpose of this paper is to present the most recent results on this experimental procedure, as obtained in our laboratory.

\section{Experimental setup}

Temperature is varied employing a helium flow cryostat and magnetic field is generated by a superconducting coil (maximum attainable field: 16 T). The electromagnetic field is generated and analysed by a Vector Network Analizer (VNA), connected to the sample by a commercial coaxial cable (cutoff frequency: 60 GHz), about 3.5 meters long (1 meter inside the cryostat and 2.5 meters outside). Such a large distance is needed in order to keep the VNA as far as possible from the magnet. On the other hand such a high length results in a heavy attenuation of the signal, especially at higher frequencies, which prevented us to perform measurements above 35 GHz and resulted in highly noisy 
measurements above 20 GHz.

The sample surface impedance is probed as a function of frequency through a Corbino disk geometry, in which the sample shortcircuits the above mentioned cable. The connection between the cable and the sample is realised through a double spring method, extensively described in \cite{toso1}.  With respect to previous realisations, a thin indium ring has been placed between the sample and the outer connector of the cable, to reduce the possible damage of the sample. However this was not sufficient to suppress a detectable contact capacitance. The capacitance effect results to be smaller for the MgB$_2$ film than for the YBCO film, suggesting that the sample surface plays a relevant role in the overall contact realisation. As a result of this spurious capacitance, data on YBCO are not reliable below 6 GHz,  while in the case of MgB$_2$ the capacitance masks data only below $2$ GHz, so that the actual frequency range over which measurements will be presented in the following of the paper is limited to $6\div 20$ GHz for YBCO and $2 \div 20$ GHz for MgB$_2$. 

The samples here considered are a thin ($d \simeq 2200$ \AA), square (l = 10 mm) YBCO film, grown by planar high oxygen pressure DC sputtering on a LaAlO$_{3}$ substrate \cite{crescitaYBCO}, and a thin ($d \simeq 1500$ \AA), square (l = 5 mm) MgB$_2$ film, grown by pulsed laser deposition with a conventional two step technique on a sapphire substrate \cite{crescitaMgB2}. 

The measured quantity is, for any magnetic field $H$ and temperature $T$, the reflection coefficient $\Gamma_m(\nu; T, H)$, measured at the VNA location. Two kinds of measurements can be made during a measuring session: the collection of a set of curves $\Gamma_{m}(\nu;T,H_{0})$ at fixed magnetic field and different temperatures, to which we will refer in the following as ``transitions'', or the collection of a set of curves $\Gamma_{m}(\nu;T_{0},H)$ at different magnetic fields and fixed temperature, which will be called ``field sweeps''. The dependencies on $\nu,T,H$ will be often omitted in the following for the sake of compactness.

\section{Measurement technique}

As discussed in the introduction, the problem issued in this paper is how to extract the surface impedance $Z_s(\nu)$ of the sample  from the reflection coefficients $\Gamma_m(\nu)$ measured at the beginning of the coaxial cable. $Z_s(\nu)$ is defined as  the ratio between transverse electric and magnetic field components at the sample surface.

It is possible to demonstrate \cite{anlage,toso1} that $Z_s(\nu)$ is related to the reflection coefficient at the sample surface $\Gamma_0(\nu)$ through the relation
\begin{equation}
\label{Zs}
Z_s(\nu) =  Z_0\frac{1+\Gamma_0(\nu)}{1-\Gamma_0(\nu)}
\end{equation}
where $Z_0$ is the characteristic impedance of the dielectric filling the cable. We estimate $Z_0 = 377 \Omega$ (vacuum impedance) being the last section of the cable not filled with the dielectric.

In principle, the values of $\Gamma_0(\nu)$ can be extracted from the measured reflection coefficient at the VNA input $\Gamma_m(\nu)$ through the standard relation \cite{Collin}
\begin{equation}
\label{gammam}
\Gamma_m = E_d+\frac{E_r\Gamma_0}{1-E_s\Gamma_0}
\end{equation}
where $E_r$, $E_s$ and $E_d$ are complex error coefficients taking into account distortions of the signal due to the transmission through the cable. They depend on frequency and on the temperature profile of the whole cable but we want to stress since now that, since the cable is constructed with non magnetic materials, the error coefficients do not depend on magnetic field. These three parameters are customarily determined by measuring the response of three known loads replacing the sample (line calibration) but in our case the experimental setup does not  allow for a full calibration of the cable since the Corbino cell is inaccessible during measuring sessions and calibration standards at the temperatures of interest in this work are not available. Nevertheless extraction of useful information from the experimental data is possible through the following two step procedure.

\subsection*{First step: definition of $\widetilde{\Gamma}_{m}(\nu)$ and  its main features}

Daily calibration of the section of the cable external to the cryostat was performed, hence from here on we will no further consider it, as if the VNA were directly connected to the internal, one-meter long, cable. This means that the cable and reflection coefficients dealt with in the following are only relative to the internal cable.

Before the beginning of the experiment we measured and stored the error coefficients $E_r$, $E_s$ and $E_d$ of the internal cable at room temperature $T_{r}$. We then define $\widetilde{\Gamma}_{m}(\nu;T)$ as:
\begin{equation}
\label{gammamtilde}
\widetilde{\Gamma}_m(\nu;T)=\frac{\Gamma_m(\nu;T)- 
E_{d}(\nu;T_{r})}{E_{r}(\nu;T_{r}) 
+E_{s}(\nu;T_{r})\left(\Gamma_m(\nu;T)-E_{d}(\nu;T_{r})\right)}
\end{equation}
As can be seen by inverting equation \ref{gammam} with respect to $\Gamma_0(\nu)$, $\widetilde{\Gamma}_{m}(\nu)$ is $\Gamma_0(\nu)$ when measured at $T=T_{r}$ and at all the other temperatures if the error coefficients were temperature independent. A relation between $\widetilde{\Gamma}_{m}(\nu;T)$ and $\Gamma_0(\nu;T)$ at $T<T_{r}$, can be found if we put $\Gamma_{m}(\nu;T)$ as obtained from equation \ref{gammam} into equation \ref{gammamtilde}: 
\begin{equation}
\label{gamma0prima}
\Gamma_{0}(T)=\frac{E_{d}(T_{r})-E_{d}(T)+\frac{E_{r}(T_{r})\widetilde{\Gamma}_{m}(T)}{1-E_{s}(T_{r})\widetilde{\Gamma}_{m}(T)}}
{E_{r}(T)+E_{s}(T)\left[E_{d}(T_{r})-E_{d}(T)+\frac{E_{r}(T_{r})\widetilde{\Gamma}_{m}(T)}{1-E_{s}(T_{r})\widetilde{\Gamma}_{m}(T)}\right]}
\end{equation}
We expect cable coefficients not to change by more than 10\% for any variation of the temperature profile in the cable. Moreover, during the room temperature calibrations, we found $|E_{d}|$ and $|E_{s}|$ to be quite generally no more than 10\% of $|E_{r}|$, which is instead of the same order of $|\widetilde{\Gamma}_{m}|\simeq 1$. This means that $|E_{d}(T_{r})-E_{d}(T)|\simeq 10^{-2}|E_{r}\widetilde{\Gamma}_{m}|$ and we can neglect it in equation \ref{gamma0prima}, obtaining after some algebra
\begin{equation}
\label{gamma0seconda}
\Gamma_{0}\simeq\frac{\frac{E_{r}(T_{r})}{E_{r}(T)}\widetilde{\Gamma}_{m}}
{1+\widetilde{\Gamma}_{m}{E_{s}(T)}\left[\Delta E_s-\Delta E_r\right]}
\end{equation}
Where $\Delta E_{s,r} = (E_{s,r}(T)-E_{s,r}(T_r))/E_{s,r}(T)$. Again $E_{s}\widetilde{\Gamma}_{m}\simeq 0.1$ and $\Delta E_{s,r}\simeq 0.1$ so that one gets, to within a few percent of precision,
\begin{equation}
\label{gamma0finale}
\Gamma_{0}\simeq\frac{E_{r}(T_{r})}{E_{r}(T)}\widetilde{\Gamma}_{m}\doteq\frac{\widetilde{\Gamma}_{m}}{\alpha(T)}
\end{equation}
We stress that the coefficient $\alpha(\nu,T)$, defined through equation \ref{gamma0finale}, depends in a non-trivial way upon the whole temperature profile of the cable so that no practical measurement of it is possible. Nevertheless, when performing measurements at low T ($T\lesssim 100$K), it is reasonable to expect that it is almost constant when varying the temperature of the sample of a few tens of Kelvin; the cable, in fact,  is thermally constrained at room temperature at the entrance of the cryostat, so that the temperature profile of the cable only slightly changes for such variations of the temperature of the sample \cite{notalphanu}. If this is the case, though $\widetilde{\Gamma}_{m}(\nu)$ is essentially different from $\Gamma_{0}(\nu)$, the ratios of $\widetilde{\Gamma}_{m}(\nu)$ are good approximations of the ratios of the corresponding $\Gamma_{0}(\nu)$. It is worth to note that equation \ref{gamma0finale} is a relation between complex quantities, implying the following couple of real equations:
$$
\left\{
\begin{array}{rcl}
   |\widetilde{\Gamma}_{m}(\nu;T)| & = &  
|\alpha(\nu;T)||\Gamma_{0}(\nu;T)| \\
   Arg[\widetilde{\Gamma}_{m}(\nu;T)] & = & 
Arg[\Gamma_{0}(\nu;T)]+Arg[\alpha(\nu;T)]
\end{array}
\right.
$$
As we will see, while $|\alpha|$ is reasonably constant over small variations of temperature, the same does not hold generally for $Arg[\alpha ]$. This poses some limit to the knowledge of  $Arg[\Gamma_{0} ]$, the consequences of which will be discussed in the following.

\subsection*{Second step: obtaining variations of $Z_{s}(\nu)$}

In this paragraph we will demonstrate how the ratio of two $\Gamma_{0}(\nu)$ determined at two different temperatures or magnetic fields is related to the corresponding difference of impedances. Some approximations are needed which we will discuss in the next section. We first consider only temperature variations and then we will extend our conclusions to the case of varying magnetic field.

Let's consider the difference between two curves of $Z_{s}(\nu)$ at two different values of the temperature, say $T_{1}$ and $T_{2}$. Making use of equation \ref{Zs}, we can write
\begin{eqnarray}
\label{deltagamma0}
\Delta Z_{s}(\nu;T_{1},T_{2})\doteq Z_{1}(\nu;T_{1}) - Z_{2}(\nu;T_{2}) &=& 
Z_{0}\left(\frac{1+\Gamma_{0}(\nu;T_{1})}{1-\Gamma_{0}(\nu;T_{1})}-\frac{1+\Gamma_{0}(\nu;T_{2})}{1-\Gamma_{0}(\nu;T_{2})}\right)\nonumber\\
              &=& Z_{0}\left(\frac{1-\frac{\Gamma_{0}(\nu;T_{1})} {\Gamma_{0}(\nu T_{2})}}{1+\frac{\Gamma_{0}(\nu;T_{1})}{\Gamma_{0}(\nu;T_{2})}-\frac{(1+\Gamma_{0}(\nu;T_{1}))(1+\Gamma_{0}(\nu;T_{2}))}{2\Gamma_{0}(\nu;T_{2})}}\right) 
\nonumber \\
           &\simeq& 
Z_0\frac{1-\Gamma_{0}(\nu;T_{1})/\Gamma_{0}(\nu;T_{2})}{1+\Gamma_{0}(\nu;T_{1})/\Gamma_{0}(\nu;T_{2})}
\end{eqnarray}
The approximation involved in the last equivalence will be extensively discussed in the following, but we stress since now that it is surely possible if $\Gamma_0(\nu;T_2)\simeq -1$. Within this limit we then found that the difference between two values of the impedance at different temperature is a function only of the ratio of the respective reflection coefficients at the sample surface. We can separate the above equation in its real and imaginary part, in order to work with real rather than complex quantities. Writing as usual $Z_{s}\doteq R_{s}+iX_{s}$, we have
\begin{equation}
    \Delta R_{s}(\nu;T_{1},T_{2})= Z_{0}\frac{1-|\frac{\Gamma_{0}(\nu;T_{1})}{\Gamma_{0}(\nu;T_{2})}|^{2}}
{1+|\frac{\Gamma_{0}(\nu;T_{1})}{\Gamma_{0}(\nu;T_{2})}|^{2}+2Re[\frac{\Gamma_{0}(\nu;T_{1})}{\Gamma_{0}(\nu;T_{2})}]}
\label{Rsprima}
\end{equation}
\begin{equation}
    \Delta X_{s}(\nu;T_{1},T_{2})= -2Z_{0}\frac{Im[\frac{\Gamma_{0}(\nu;T_{1})}{\Gamma_{0}(\nu;T_{2})}]}{1+|\frac{\Gamma_{0}(\nu;T_{1})}{\Gamma_{0}(\nu;T_{2})}|^{2}+2Re[\frac{\Gamma_{0}(\nu;T_{1})}{\Gamma_{0}(\nu;T_{2})}]}
\label{Xsprima}
\end{equation}
Now we come to the last approximation. We define $\Delta\phi_{0}(\nu)\doteq Arg[\Gamma_{0}(\nu;T_{1})]-Arg[\Gamma_{0}(\nu;T_{2})]$ as the phase variation between the two reflection coefficients; if $\Delta\phi_{0}(\nu)$ is small (for each frequency $(\nu)$ of interest) then the approximations $Re\frac{\Gamma_{0}(\nu;T_{1})}{\Gamma_{0}(\nu;T_{2})}\simeq \left|\frac{\Gamma_{0}(\nu;T_{1})}{\Gamma_{0}(\nu;T_{2})}\right|$ and $Im\frac{\Gamma_{0}(\nu;T_{1})}{\Gamma_{0}(\nu;T_{2})}\simeq \left|\frac{\Gamma_{0}(\nu;T_{1})}{\Gamma_{0}(\nu;T_{2})}\right|\Delta\phi_{0}$ 
hold. In this case equations \ref{Rsprima} and \ref{Xsprima} can be approximated as
\begin{equation}
    \Delta R_{s}(\nu;T_{1},T_{2})\simeq Z_{0}\frac{1-|\frac{\Gamma_{0}(\nu;T_{1})}{\Gamma_{0}(\nu;T_{2})}|}{1+|\frac{\Gamma_{0}(\nu;T_{1})}{\Gamma_{0}(\nu;T_{2})}|}
    \label{Rs0}
\end{equation}
\begin{equation}
    \Delta X_{s}(\nu;T_{1},T_{2})\simeq 
-2Z_{0}\frac{|\frac{\Gamma_{0}(\nu;T_{1})}{\Gamma_{0}(\nu;T_{2})}|}{(1+|\frac{\Gamma_{0}(\nu;T_{1})}{\Gamma_{0}(\nu;T_{2})}|) 
^{2}}\Delta\phi_{0}
\label{Xs0}
\end{equation}
where we stress that equation \ref{Rs0} depends only on the ratio of the moduli of the reflection coefficients and not upon the  phase variations. In this way the determination of $\Delta R_{s}$ requires information about $|\Gamma_0|$ only, while $\Delta X_s$ can be obtained only through the knowledge of both $|\Gamma_0|$ and $\Delta \phi_0$.

Putting the results of equation \ref{gamma0finale} into these formulae we get
\begin{equation}
    \Delta R_{s}(\nu;T_{1},T_{2})\simeq Z_{0}\frac{1-\frac{|\alpha(\nu;T_{2})|}{|\alpha(\nu;T_{1})|}\frac{|\widetilde{\Gamma}_{m}(\nu;T_{1})|}{|\widetilde{\Gamma}_{m}(\nu;T_{2})|}}{1+\frac{|\alpha(\nu;T_{2})|}{|\alpha(\nu;T_{1})|}\frac{|\widetilde{\Gamma}_{m}(\nu;T_{1})|}{|\widetilde{\Gamma}_{m}(\nu;T_{2})|}}
    \label{Rsm}
\end{equation}
\begin{equation}
    \Delta X_{s}(\nu;T_{1},T_{2})\simeq -2Z_{0}\frac{\frac{|\alpha(\nu;T_{2})|}{|\alpha(\nu;T_{1})|}\frac{|\widetilde{\Gamma}_{m}(\nu;T_{1})|} 
{|\widetilde{\Gamma}_{m}(\nu;T_{2})|}}{(1+\frac{|\alpha(\nu;T_{2})|}{|\alpha(\nu;T_{1})|}\frac{|\widetilde{\Gamma}_{m}(\nu;T_{1})|}
{|\widetilde{\Gamma}_{m}(\nu;T_{2})|})^{2}}(\Delta\phi_{m}-\Delta\phi_{\alpha})
\label{Xsm}
\end{equation}
If we have $|\alpha(\nu;T_{1})|\simeq |\alpha(\nu;T_{2})|$, so that $\frac{|\Gamma_{0}(\nu;T_{1})|}{|\Gamma_{0}(\nu;T_{2})|}\simeq\frac{|\widetilde{\Gamma}_{m}(\nu;T_{1})|}{|\widetilde{\Gamma}_{m}(\nu;T_{2})|}$, equation \ref{Rs0} can be finally rewritten as
\begin{equation}
    \label{Rsfinale}
    \Delta R_{s}(\nu;T)\simeq Z_{0}\frac{1-\frac{|\widetilde{\Gamma}_{m}(\nu;T_{1})|}{|\widetilde{\Gamma}_{m}(\nu;T_{2})|}}{1+\frac{|\widetilde{\Gamma}_{m}(\nu;T_{1})|}{|\widetilde{\Gamma}_{m}(\nu;T_{2})|}}
\end{equation}
If we also have $\Delta\phi_{\alpha}\ll\Delta\phi_{0}$, so that $\Delta\phi_{0}\simeq\Delta\phi_{m}$, then equation \ref{Xs0} can be approximated with
\begin{equation}
    \label{Xsfinale}
    \Delta X_{s}(\nu;T)\simeq  -2Z_{0}\frac{\frac{|\widetilde{\Gamma}_{m}(\nu;T_{1})|}{|\widetilde{\Gamma}_{m}(\nu;T_{2})|}}{(1+\frac{|\widetilde{\Gamma}_{m}(\nu;T_{1})|}{|\widetilde{\Gamma}_{m}(\nu;T_{2})|}) ^{2}}\Delta\phi_{m}
\end{equation}

Formulae \ref{Rsfinale} and \ref{Xsfinale} are the main result of this paper. They relate the variation of impedance between two different temperatures to the measured quantities. Through these formulae, in fact, once a temperature $T_{ref}$ at which $\Gamma_{0}\simeq -1$ is fixed, the differences $\Delta Z_{s}(\nu;T)\doteq Z_{s}(\nu;T)-Z_{s}(\nu;T_{ref})$ for each temperature $T$ can be obtained. Moreover, once all the $\Delta Z_s$ are known, if a temperature $T^{*}$ exists for which the curve $Z_{s}(\nu;T^{*})$ is theoretically predictable or experimentally known, absolute values of $Z_{s}(\nu;T)$ can be extracted.

The procedure can be easily extended to the case of measurements as a function of the applied magnetic field at a fixed temperature $T_0$ (field sweeps). In this case, $\Delta Z_s =\Delta Z_s(\nu ;H) \doteq Z_s(\nu;H,T_0) - Z_s(\nu;H_{ref},T_0)$ and the values of $\Delta R_s(\nu ,H)$ and $\Delta X_s(\nu ,H)$ are obtained with the same expressions as in equations \ref{Rsfinale} and \ref{Xsfinale} by replacing the temperature dependence with the magnetic field dependence. With respect to measurements as a function of temperature, there is the advantage that the response of the cable is not expected to vary by applying an external magnetic field. This implies that the coefficient $\alpha(\nu)$  remains constant among measurements at different magnetic fields, provided that the whole cable has reached its equilibrium temperature profile. The result is that the condition $\alpha(\nu;H)=\alpha(\nu;H_{ref})$ needed to obtain equations \ref{Rsfinale} and \ref{Xsfinale} from \ref{Rsm} and \ref{Xsm} is exactly, instead of approximately, satisfied. 

\subsection*{Check of the validity of the approximations}

In this paragraph we discuss the validity of the approximations introduced in the previous section to come to equations \ref{Rsfinale} and \ref{Xsfinale}. Two of these approximations deal with hypotheses on $\Gamma_0$ (markedly, that it exists a pair of values \{H,T\} for which $\Gamma_0(H,T,\nu)\simeq -1$ and that the variation of $Arg[\Gamma_0]$ as a function of $T$ and $H$ are small) while the other two approximations deal with the variation as a function of $T$ and $H$ of the complex parameter $\alpha$ relating $\Gamma_0$ with $\widetilde\Gamma_m$. This task will be accomplished by putting together information that can be obtained from DC and microwave measurements, as well as from theoretical predictions.

\begin{figure}
\centerline{\psfig{figure=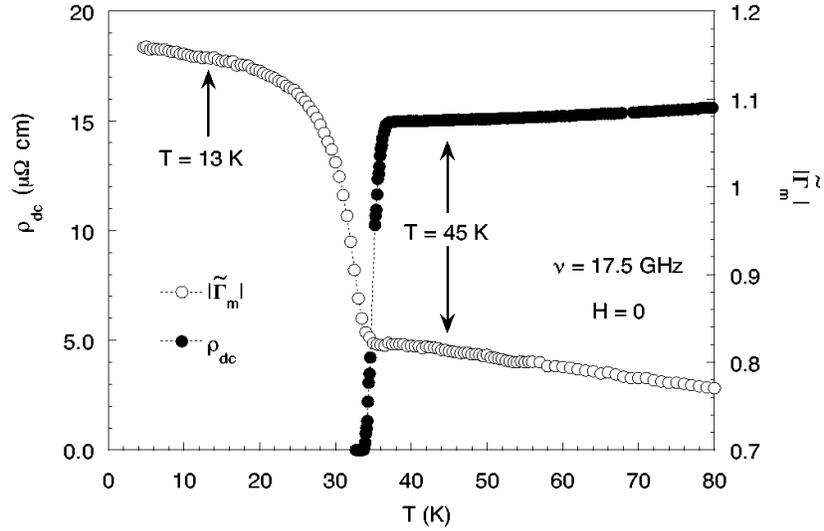,height=7cm,width=11cm}}
\caption{Right-hand scale, empty circles: modulus of $\widetilde{\Gamma}_{m}(\nu,T)$ at fixed frequency $\nu =17.5$ GHz as a function of temperature and zero magnetic field. Left-hand scale, full circles: DC resistivity of the same sample at zero field.}
\label{DC+Gamma}
\end{figure}

As a first step, we notice that from the resistivity value obtained from DC measurements (see figure \ref{DC+Gamma}) we can infer that the thin film approximation \cite{silva} is valid just above the transition temperature. We will then assume that for all temperatures one may write $Z_s \simeq \rho/d$.
We then focus on the assumptions related to $\Gamma_0$. The first assumption is certainly verified at low enough temperatures and zero applied magnetic field, since both $R_s$ and $X_s$ are expected to become much smaller than $Z_0$ for $T\ll T_c$, so that $\Gamma_0 = (Z_s-Z_0)/(Z_s+Z_0) \simeq -1$. We will come later on the criteria used to choose the value of $T_{ref}$, and to the choice of $H_{ref}$ and $T_{ref}$ for the measurements as a function of the applied magnetic field.

\begin{figure}
\centerline{\psfig{figure=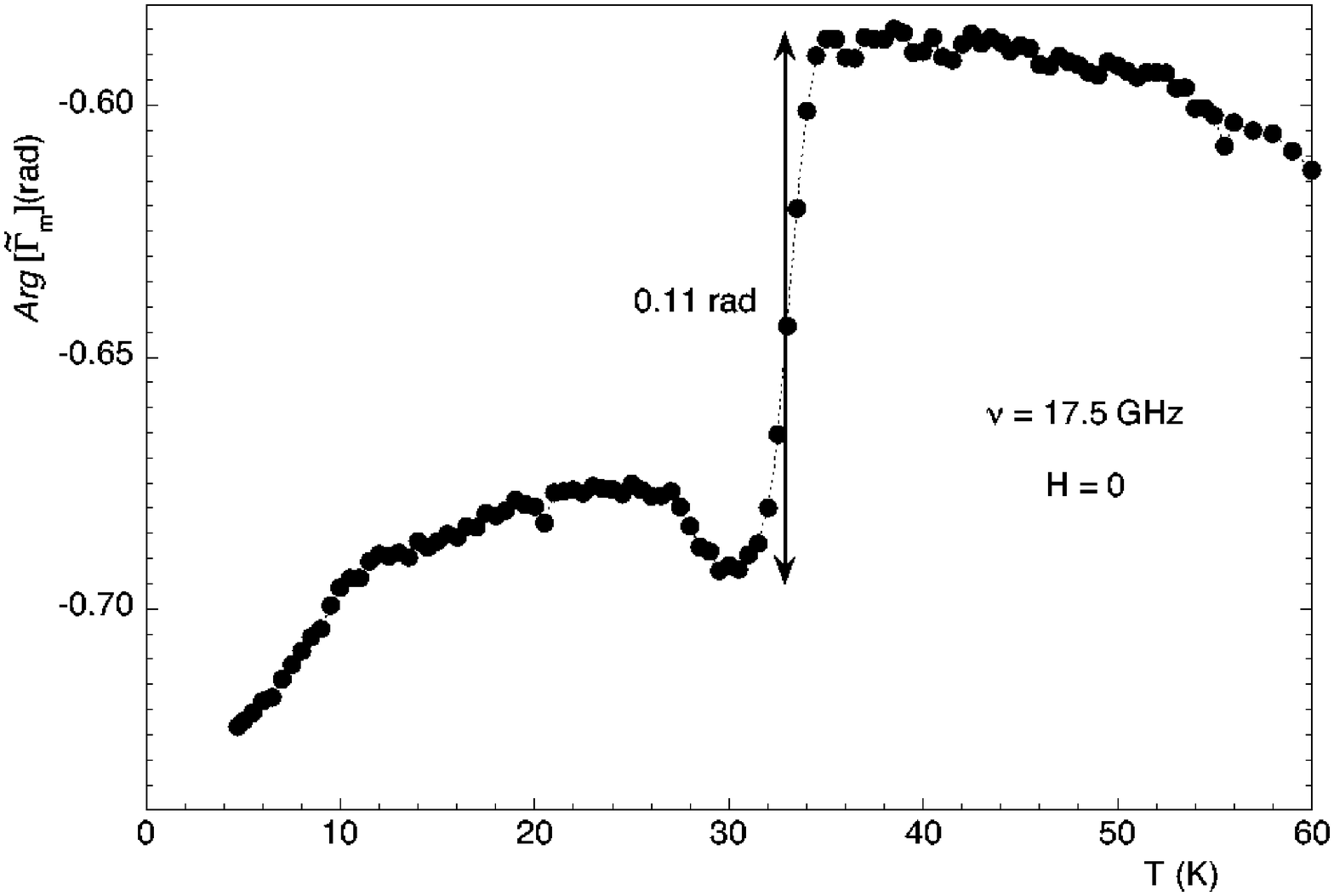,height=7cm,width=11.5cm}}
\caption{Phase of $\widetilde{\Gamma}_{m}(\nu,T)$ at fixed frequency $\nu =17.5$ GHz as a function of temperature and fixed zero feld for MgB$_{2}$.}
\label{phase}
\end{figure}

The second approximation can be verified by measuring the temperature variations of the phase of the measured $\widetilde\Gamma_m$. In figure \ref{phase} we report the behaviour of $Arg[\widetilde\Gamma_m]$ as a function of $T$ for $H=0$ T and $\nu = 17.5$ GHz. The overall behaviour is composed by a rather sharp peak, located in a $T$ range just below $T_c$, over a smoothly $T$ dependent background. Since the phase of $\Gamma_0$ is expected to be constant, as a function of $T$, both above $T_c$ (where $Z_s$ and thus $\Gamma_0$ are real) and sufficiently below $T_c$ (where, as discussed above, $Arg[\Gamma_0]\simeq \pi$), it is natural to ascribe the smooth background to variations of $Arg[\alpha]$ and the peak to variations of $\Gamma_0$. The peak of $Arg[\Gamma_0]$ can be easily understood on general grounds: in fact, $Arg[\Gamma_0]\neq 0$ only if $X_s$ is comparable with $Z_0$, which is true only at temperatures slightly below $T_c$. Similar results are obtained also when it is present an applied magnetic field. From the measured values, one gets $\Delta\phi_0 \leq 0.1$ rad , so that the approximation $\Delta\phi_0 \ll 1$ is reasonably fulfilled for all temperatures and fields.

We now come to the approximations dealing with the variations of $\alpha$, starting with the approximation $|\alpha| \simeq const$. In figure \ref{DC+Gamma} we plot the measured DC resistivity of the MgB$_2$ sample, together with the measured $|\widetilde\Gamma_m|$ at 17.5 GHz. Since $\rho$ is expected to be real and independent on frequency above $T_c$, one has $Z_s$ real in this temperature range, resulting in $\Gamma_0$ being real and independent on frequency. Further, since $\rho_{DC}$ is almost independent on $T$ for $35$K$ < T < 45 $K, we may conclude that $\Gamma_0$ is also almost independent on $T$ just above $T_c$. The measured $|\widetilde\Gamma_m|$ is approximately constant, as a function of temperature, between $45$ K and $35$ K, as well as below $13$ K. This is consistent with the above discussion: above $T_c$, one has $\Gamma_0$ almost independent on $T$, and sufficiently below $T_c$ one has $|\Gamma_0|\simeq 1$, again independent on $T$. The small variation as a function of $T$ can then be ascribed to variations of $|\alpha|$ with $T$, which turn out to be much smaller than the variations of $|\Gamma_0|$, as required. It is worth to stress that at high frequency one has $R_s\simeq 0$ only for $T\ll T_c$, so that it is not surprising that the condition  $\widetilde\Gamma_m \simeq -1$ is reached only for $T\simeq 13$K.  As noticed before, this fact, together with the validity of the assumptions on $\Gamma_0$ previously verified, guarantees that it is possible to obtain the values of $R_s$ from the measured $\Gamma_m$. 

The last approximation to be verified is the one related to variations of $Arg[\alpha]$. It is evident from figure \ref{phase} that the peak of $Arg[\Gamma_0]$ is of the same order of magnitude as the variations of the smooth background. This means that the condition $\Delta\phi_{\alpha}\ll\Delta\phi_{0}$ does not hold and we cannot make use of equation \ref{Xsfinale} to get sample reactance from measurements as a function of temperature. As noticed before, this limitation is intrinsically eliminated when performing measurements at fixed temperature as a function of the applied magnetic field (field sweeps), that allow for a determination of both the real and the imaginary part of $\Delta Z_s$.

\subsection*{Obtaining the impedances}

Once the validity of the approximations has been discussed, we turn to the description of the practical method employed for the determination of the impedance. This purpose is pursued through extensive use of equations \ref{Rsfinale} and \ref{Xsfinale}. We start with the zero field case. First of all it is necessary to choose the reference temperature $T_{ref}$ with respect to which we calculate the variation of the impedance. This choice is dictated by the necessity to satisfy the approximation involved in equation \ref{deltagamma0}.  As we saw in the previous paragraph this approximation is surely valid for the zero field transition when the choice $T_{ref}\ll T_{c}$ is made. Operatively the condition $T\ll T_{c}$ is considered to be reached when the curves $|\widetilde\Gamma_m (\nu;T,0)|$  approach a constant value as a function of temperature (in figure \ref{DC+Gamma}, for example, below 13 K). The temperature $T_{ref}$ is then chosen as low as possible considering that we need at the same time $\alpha\simeq const$ in the temperature interval $[T_{ref},T_{c}]$. We then choose $T_{ref}=4$ K for MgB$_{2}$ and $T_{ref}=70$ K for YBCO \cite{Tc-Tref}. 

Once we have all the differences $\Delta R_s(\nu;T,0)=R_{s}(\nu;T,0)-R_{s}(\nu;T_{ref},0)$ we can obtain the absolute values $R_{s}(\nu;T,0)$ if one of the curves $R_{s}(\nu;T^{*},0)$ is known. Again, as we discussed in the previous paragraph, at $T\ll T_{c}$ $R_{s}\simeq 0$ so that in this case the extraction of the absolute values is trivial  because, with the choice of $T_{ref}\ll T_{c}$, $\Delta 
R_s(\nu;T,0)=R_{s}(\nu;T,0)$.

For whet concerns the field sweeps measurements we chose $H_{ref}=0$ for all sweeps. In this case, when the temperature $T_{0}$ at which the field sweep is performed is close to $T_{c}$, the condition $\Gamma_0(\nu;T_{0},H_{ref})\simeq -1$ is not fully satisfied. Nevertheless, by using the values of $R_{s}(\nu;T,0)$ obtained from the zero field transition, we can estimate in no more than 5\% the error made with the approximation contained in equation \ref{deltagamma0}.

Absolute values of $R_{s}$ can again be extracted for field sweeps measurements using the curve $R_{s}(\nu;T_{0},0)$ obtained from the zero field transition. Since, by choosing $H_{ref}=0$, all the curves $\Delta R_{s}$ are variations with respect to $H=0$, we add to all the curves of the field sweep the curve $R_{s}(\nu;T_{0}H=0)$, obtained from the zero field transition. In this way we transform the set of curves $\Delta R_{s}(\nu;T_{0},H)$ into a set of curves $R_{s}(\nu;T_{0},H)$.

To obtain absolute values of $X_{s}$ we cannot make use of the zero field transition because we recall that $X_{s}$ cannot be extracted from transitions. However we can exploit the fact that  above the upper critical field, where the system is in the normal state, $X_s$ is expected to vanish. Then, by subtracting to each curve $\Delta X_s(\nu;T_{0},H)$ the value of $\Delta X_s(\nu;T_{0},H^{*})$, with $H^*>H_{c2}$, one gets  the absolute values of $X_s(\nu;T_{0},H)$. In order to check for which values of $H$ the condition $H>H_{c2}$ is satisfied, we analyse the curves $\Delta X_s(\nu;T_{0},H)$ in a similar way as was done  to determine $T_{ref}$ in the zero field transition: it can be noticed, in fact, that $\Delta X_s(\nu;T_{0},H)$ approaches a constant value at high magnetic fields, so that we choose $H^{*}$ among the fields at which $\Delta X_s(\nu;T_{0},H)\simeq const$. 

This latter procedure is possible only if the observed $\Delta X_s(\nu;T_{0},H)$ reaches a constant value at high fields, that is if the applied magnetic field reaches a value above the upper critical field. This can easily be obtained MgB$_2$, whose upper critical field is around 10 T even at low temperature, while it is not possible for YBCO, whose upper critical field exceeds our experimental limits a few Kelvin below $T_c$. The results will then be presented as $R_s(\nu;T,H)$ and $X_s(\nu;T,H)$ for MgB$_2$, while for YBCO we can report only $R_s(\nu;T,H)$ and $\Delta X_s(\nu;T,H)$.

\section{Results}

\begin{figure}
\centerline{\psfig{figure=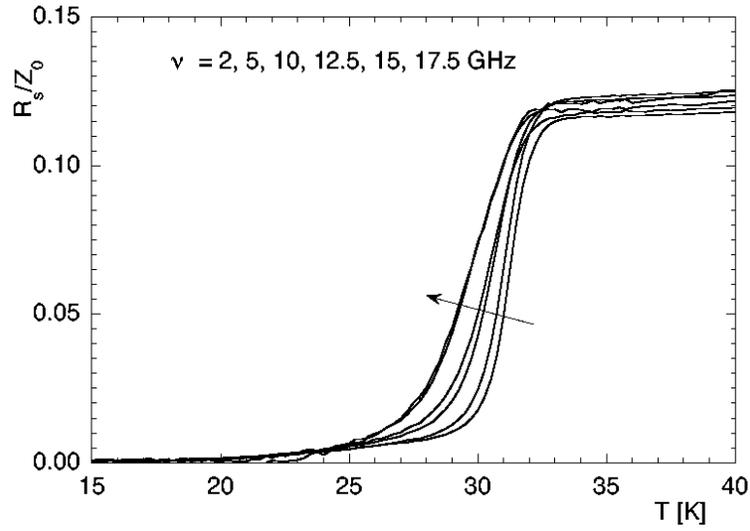,height=7.cm,width=10cm,clip=,angle=0.}}
\caption{Measured resistivities as a function of temperature, at several fixed frequencies. Note the widening of the transition, as the frequency is increased.}
\label{R1(T)_MgB2}
\end{figure}

\begin{figure}
\centerline{\psfig{figure=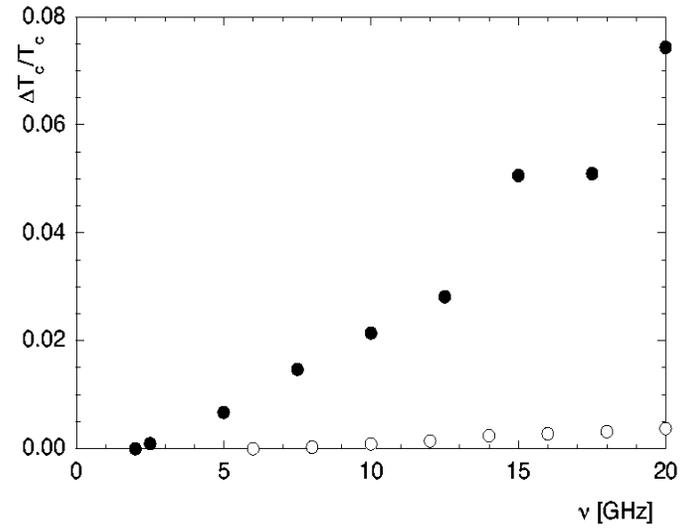,height=7.cm,width=9.cm,clip=,angle=0.}}
\caption{Widening of the transition as a function of frequency, for both YBCO (open symbols) and MgB$_2$ (filled symbols) (see text for the definition of $\Delta T_c/T_c$)}
\label{Delta_Tc}
\end{figure}

In the following, we report some examples of the measuring procedure for both MgB$_2$ and YBCO. We first present the results of the measurement at zero magnetic field as a function of temperature. In fig. \ref{R1(T)_MgB2} we report the real part of $Z_s(T)$ in MgB$_2$ at different frequencies. The transition widens with increasing frequency. To have a more quantitative description of the widening, one can define $T_{mp}(\nu)$ as the temperature where the resistivity, at the given frequency $\nu$, reaches one half of the value in the normal state, and  $\Delta T_c (\nu) = T_{mp}(\nu) - T_{mp}(\nu_{min})$. The behaviour of $\Delta T_c(\nu)$, normalised to the value of $T_c$ at $\nu_{min} = 2$ GHz is reported in fig. \ref{Delta_Tc} (filled circles).

\begin{figure}
\centerline{\psfig{figure=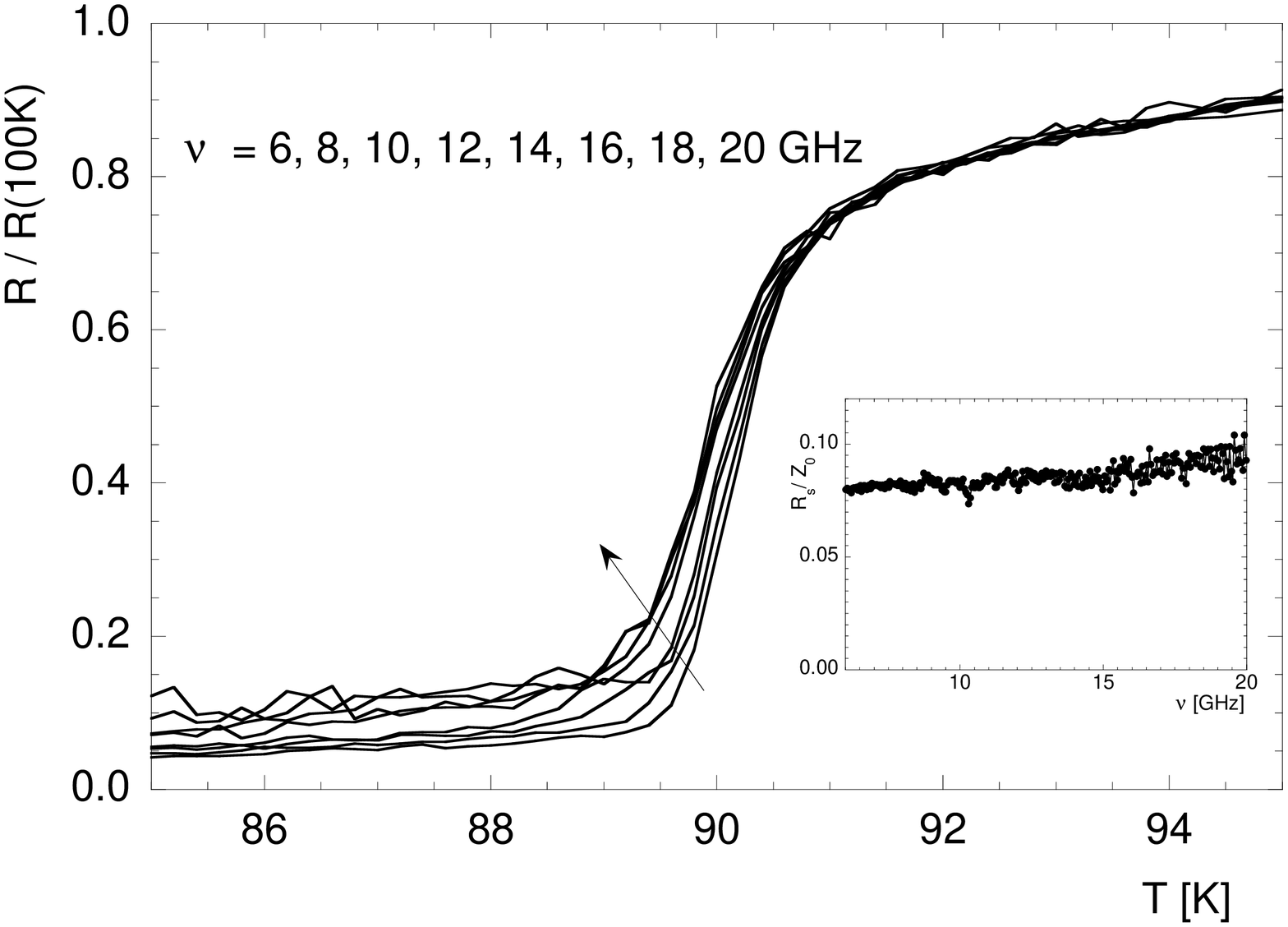,height=7.cm,width=9.cm,clip=,angle=0.}}
\caption{Measured resistivities as a function of temperature for the YBCO film. The resistivities has been normalized to their value obtained at 100 K to reduce the effect of the variation of the cable parameters between 70 and 100 K. The behaviour of $R_s$ at 100 K is reported in the insert.}
\label{R1(T)_YBCO}
\end{figure}

The same analysis is presented in fig.\ref{R1(T)_YBCO}, for YBCO where curves of $R_s/R_s(100$K) are presented as a function of $T$ at various frequencies. The real part of the resistivity remains almost zero up to $\simeq 88\;$K, then grows at $T_c$ rapidly reaching the normal state value. As in the case of MgB$_2$, the transition widens as the frequency is increased. However, the widening is relatively small compared to MgB$_2$ and the upper part of the transition is almost unaffected by the variation of the frequency, being only the very last part of it severely dependent on frequency. The values of $\Delta T_c / T_c$ (with $\nu_{min} = 6$ GHz) are reported in fig. \ref{Delta_Tc}  for a quantitative comparison with the values obtained  for MgB$_2$.

\begin{figure}
\centerline{\psfig{figure=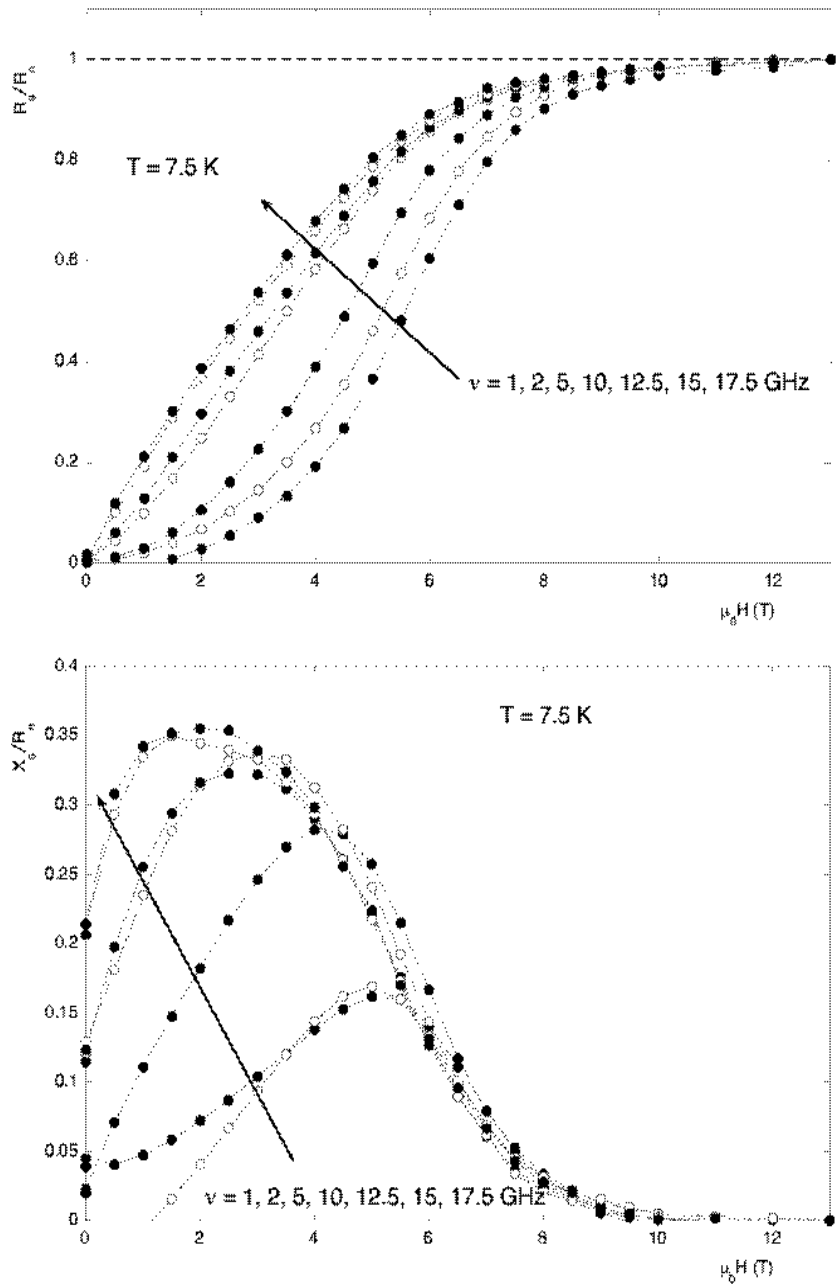,height=17.cm,width=11.cm,clip=,angle=0.}}
\caption{Measurement of $R_s$ (upper panel) and $X_s$ (lower panel) in MgB$_2$ at $T=7.5$K and various frequencies, as a function of the applied magnetic field.}
\label{MgB2_7.5K}
\end{figure}

\begin{figure}
\centerline{\psfig{figure=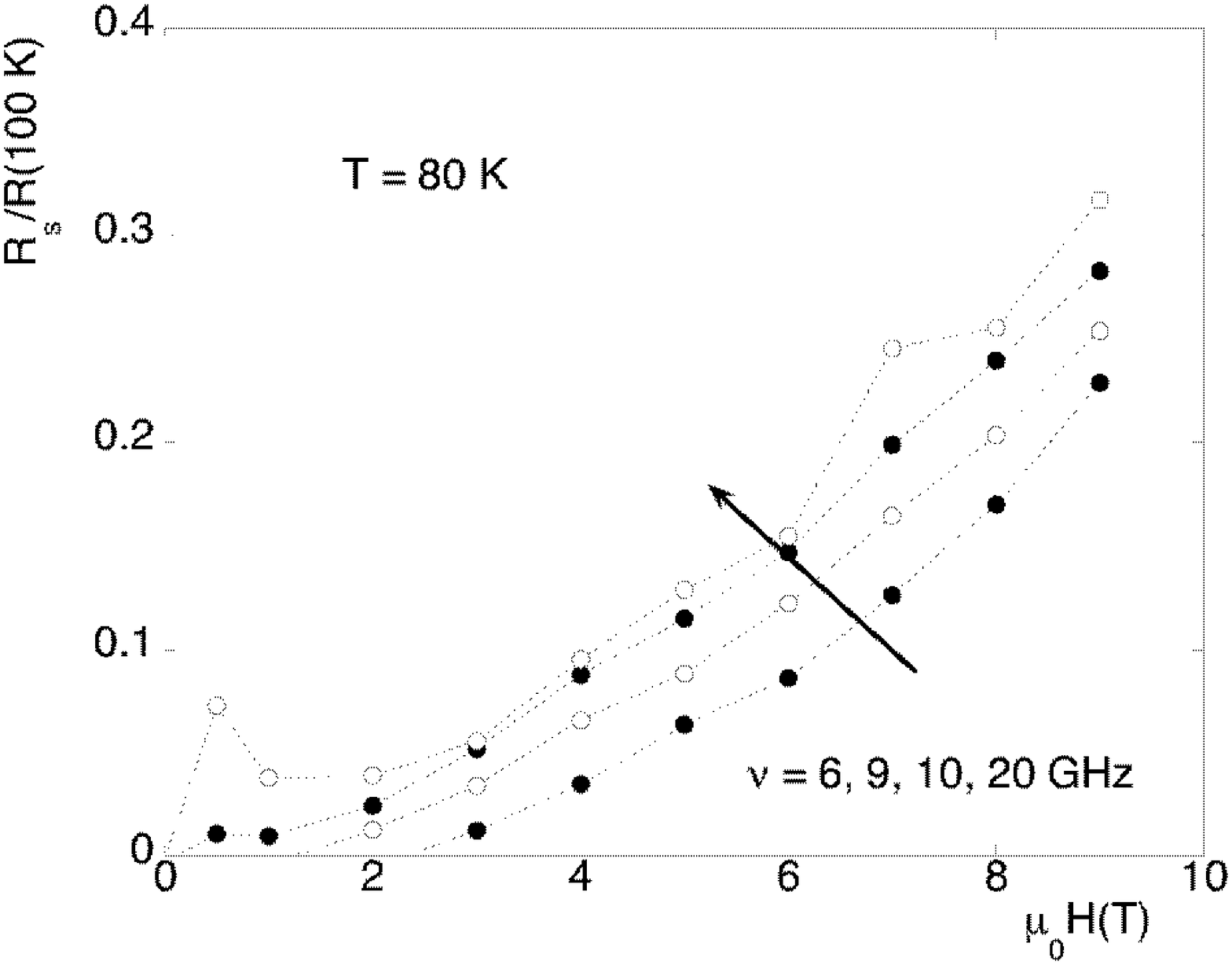,height=7.cm,width=9.cm,clip=,angle=0.}}
\caption{Measurement of $R_s$  in YBCO at $T=80$K and various frequencies, as a function of the applied magnetic field.}
\label{YBCO-ramp}
\end{figure}

We now come to the results as a function of magnetic field, at fixed frequency. Again, we first present results on MgB$_2$. In figure \ref{MgB2_7.5K} we report measurements of $R_s/R_n$ and $X_s/R_n$ (where $R_n$ is the resistivity at the highest field) at 7.5 K and different frequencies, as a function of the applied magnetic field. Both $R_s$ (upper panel) and $X_s$ (lower panel) do not show appreciable variations as a function of field above $\mu_0 H \simeq 9$ T, indicating that the upper critical field is reasonably below that value. The assumption that at 13 T the material is in its normal state, so that $R_s(\nu;13$T$) = R_n$ and $X_s(\nu;13$T$)=0$ is then fully justified. Finally, we present in figure \ref{YBCO-ramp} the real part of the resistivity in YBCO at $T = 80$ K, at various frequencies and as a function of the applied magnetic field.

\section{Conclusions}

We collected reflection coefficient measurements of  superconducting thin films as a function of frequency through a Corbino disk technique. We developed a measurement technique by which it is possible to obtain the impedance from the reflection coefficients once that some conditions are verified. We extensively discuss these conditions and how we can conclude which of them are verified. The relative curves at fixed frequencies are shown, together with a brief discussion of the data.

\newpage

\begin{figure}
\centerline{\psfig{figure=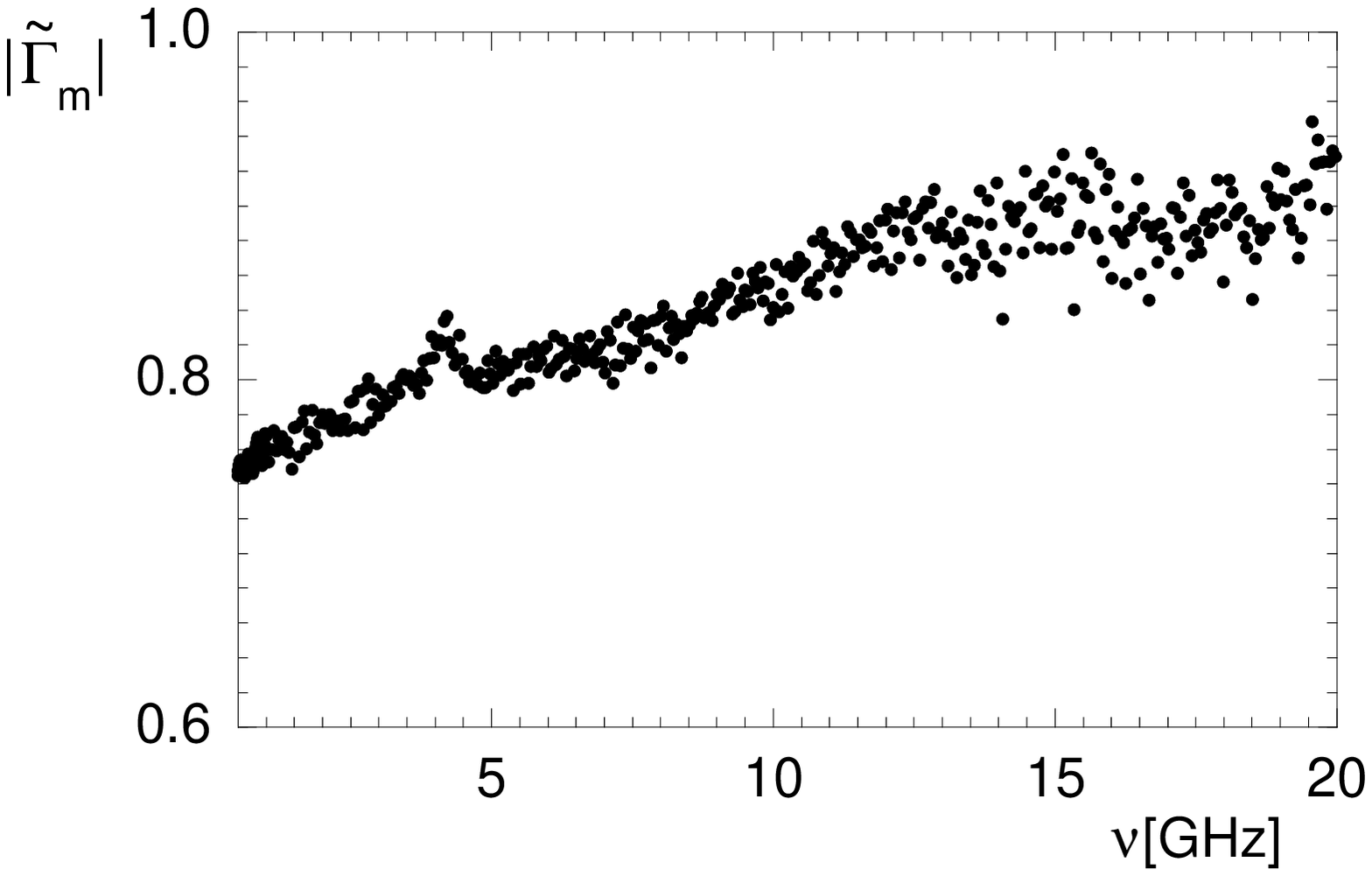,height=7.cm,width=10.5cm,clip=,angle=0.}}
\caption{Typical curve of $|\widetilde{\Gamma}_{m}|(\nu;T)$ at fixed temperature T=40K as a function of frequency for MgB$_{2}$. Since $|\Gamma_{0}|$ is reasonably constant with frequency at this temperature we attribute the frequency dependence of $|\widetilde{\Gamma}_{m}|(\nu;40K)$ to a frequency dependence in 
$|\alpha(\nu;40K)|$.}
\label{Gmtilde(nu)}
\end{figure}

\begin{thebibliography}{1}

\bibitem{anlage} D.H.Wu, J.C.Booth, S.M.Anlage, \textit{Phys. Rev. Lett.}, {\bf 75}, 525 (1995)

\bibitem{toso1} N.Tosoratti, R.Fastampa, M.Giura, V. Lenzi, S.Sarti, E.Silva, \textit{Int. J. of Mod. Phys.} B, {\bf 14}, 2926 (2000). Note that using the notation of that paper $Z_0 = 1/Y_0$.

\bibitem{crescitaYBCO} C.Beneduce, F.Bobba, M.Boffa, A.M.Cucolo, A.Andreone, C.Aruta, M.Iavarone, F.Palomba, G.Pica, M.Salluzzo, R.Vaglio, \textit{Int. J. of Mod. Phys. B}, {\bf 13}, 1333(1999)

\bibitem{crescitaMgB2} V.Ferrando, S.Amoruso, E.Bellingeri, R.Bruzzese, P.Manfrinetti, D.Marr\`e, N.Spinelli, R.Velotta, X.Wang, C.Ferdeghini, \textit{Supercond. Sci. Technol.} {\bf 16}, 241(2003)

\bibitem{silva} E.Silva, M.Lanucara, R.Marcon, \textit{Supercond.Sci.Technol.} {\bf 9}, 934 (1996)

\bibitem{Collin} R.E.Collin, \textit{ Foundation for microwave engineering} McGraw-Hill (1966)

\bibitem{notalphanu} We can also roughly estimate the frequency dependence of $\alpha(\nu;T)$ considering that it is the ratio between two transmission coefficients at different temperatures. Since $E_{r}(\nu;T)$ takes into account the attenuation of the field due to losses through the dielectric and on the cable conductors, it decreases with increasing frequency and temperature. We can then expect that, if $T<T_{r}$, $|\alpha(\nu;T)|>1$ and it will increase with frequency. This is exactly the behaviour measured, as can be seen in figure \ref{Gmtilde(nu)}.

\bibitem{Tc-Tref} Since the condition $\Gamma_0\simeq -1$ is more and more satisfied as $T\ll T_c$, one might be tempted to use $T=4$ K for both YBCO and MgB$_2$. However, an important requirement is that $\alpha$ be constant over the whole $T$ range between $T_{ref}$ and the temperature $T$ of interest. It is the necessary to limit the overall temperature range explored, in order to minimise the effects of temperature variations of $\alpha$.

\end{thebibliography}
\end{document}